\documentclass{iau} 
\usepackage{graphicx}

\title[Chemical tagging in the halo] 
{Rediscovering the origins of the stellar halo with chemical tagging}

\author[Sarah L Martell]   
{Sarah L Martell$^1$}

\affiliation{$^1$School of Physics \\ University of New South Wales \\ 2052 Sydney, Australia \\ email: {\tt s.martell@unsw.edu.au} \\[\affilskip]}

\pubyear{2017}
\volume{134}  
\jname{Rediscovering the Milky Way}
\editors{C. Chiappini, I. Minchev. E. Starkenburg \& M. Valentini, eds.}
\begin{document}

\maketitle

\begin{abstract}
The Galactic halo has a complex assembly history, which can be seen in its wealth of kinematic and chemical substructure. Globular clusters lose stars through tidal interactions with the Galaxy and cluster evaporation processes, meaning that they are inevitably a source of halo stars. These ``migrants'' from globular clusters can be recognized in the halo field by the characteristic light element abundance anticorrelations that are commonly observed only in globular cluster stars, and the number of halo stars that can be chemically tagged to globular clusters can be used to place limits on the formation pathways of those clusters.

\end{abstract}

\firstsection 
\section{Introduction}
The Galactic halo has a complex assembly history. This can be seen in the amount of substructure that has not yet settled into a smooth configuration, both spatially (e.g., \cite[Belokurov et al. 2006a]{BZ06}) and kinematically (e.g., \cite[Helmi et al. 2017]{HV17}). The chemodynamic complexity of the halo is also a prediction of hierarchical structure formation (e.g., \cite[Bullock \& Johnston 2005]{BJ05}; \cite[Tissera et al. 2014]{TB14}). 

However, the halo is not formed exclusively through accretion: because globular clusters lose stars in the course of their normal evolution, they must contribute stars to the halo field at some level. Some of the stars that have migrated from globular clusters to the halo can be recognized by the presence of a characteristic light-element abundance pattern that is seen in all globular clusters and appears to be unique to star formation in a globular cluster environment.

This paper discusses the processes by which globular cluster stars escape into the halo field (Section \ref{sec:intro}), the abundance signatures that can be used to identify their origins (Section \ref{sec:abunds}), an overview of previous and ongoing work to identify these stars in spectroscopic data sets (Section \ref{sec:tagging}), and finally a discussion of the next steps, both observational and theoretical, that are most important to expand our understanding of globular clusters and their role in the formation of the Milky Way (Section \ref{sec:interpretation}).

\section{Globular cluster contributions to halo assembly}\label{sec:intro}
There are a number of processes by which stars escape from globular clusters. Tidal interactions with the Galaxy create extended, narrow tails, (e.g., Palomar 5, \cite[Odenkirchen et al. 2001]{OG01}) which can extend for many degrees across the sky. Mass segregation acts to shift the lowest-mass stars into the least bound region of a cluster, making those stars more likely to escape (e.g., \cite[Balbinot \& Gieles 2017]{BG17}), and two-body relaxation feeds stars into the high end of the cluster velocity distribution, driving evaporation. The relative importance of the various mass-loss processes depends on each cluster's mass, concentration and orbit, as shown in the ``vital diagram'' of \cite{GO97}. High cluster concentrations increase the effectiveness of evaporation but decrease the effectiveness of disk and bulge shocking, and clusters on small-radius orbits are more susceptible to shocking because of their more frequent perigalacticon passages. Overall we expect each globular cluster to have lost roughly half of its stars over the past 12 Gyr (e.g., \cite[Fall \& Zhang 2001]{FZ01}).

We can see this mass loss in action at the present day when it is spatially coherent (e.g., \cite[Belokurov et al. 2006b]{BE06}) or when the stars have enough observable properties (e.g., position, radial velocity, proper motion, metallicity) in common with their cluster of origin (e.g., \cite[Navin et al. 2015]{NM15}; \cite[2016]{NM16}). However, large-scale tidal features are rare at the present time (e.g., \cite[Jordi \& Grebel 2010]{JG10}). This is partly because the stars in tidal tails are preferentially low-mass and therefore too faint for standard matched-filter methods (\cite[Balbinot \& Gieles 2017]{BG17}), and partly because the dramatic mass loss needed to produce and sustain large tidal tails tends to be very disruptive to a cluster, making it necessarily a short-lived phase. Within a few orbital timescales, tidal tails and extratidal stars will lose their spatial coherence and become anonymous members of the halo field, and clusters undergoing strong mass loss will dissociate entirely.
 
\section{Globular cluster abundance signatures}\label{sec:abunds}
Globular cluster stars do not blend perfectly into the halo field, however. Globular clusters exhibit anticorrelations in their light-element abundances that are not common in other components of the Milky Way. While roughly half of the stars in each cluster have an ordinary scaled-Solar abundance pattern, the other half have depletions in carbon, oxygen and magnesium abundances, and enhancements in nitrogen, sodium and aluminium abundances (e.g., \cite[Kraft 1979]{K79}; \cite[Norris \& Freeman 1979]{NF79}; and many references since). 


Globular cluster stars with this characteristic abundance pattern that escape from globular clusters will continue to be recognizable as having originated within a globular cluster, even as their orbits mix into the halo field. This is a kind of chemical tagging (e.g., \cite[Freeman \& Bland-Hawthorn 2002]{FBH02}) at the population level, in which stars are marked as having a common origin not by precisely matching abundances but by key abundance signatures that uniquely identify their formation sites. Globular clusters are an ideal environment for this kind of chemical tagging, since they appear to be the only star formation site in the Galaxy that imparts this anticorrelated light-element abundance pattern.

\section{Population-level chemical tagging}\label{sec:tagging}
A first demonstration of this method for identifying globular cluster migrants in the field was carried out by \cite{MG10}. That study used roughly 2000 low-metallicity halo giants from the SDSS-II SEGUE survey. Using the CN and CH band strengths, which are proxies for the abundances of nitrogen and carbon, respectively, the authors found that 2.5\% of halo giants had both depleted carbon and enhanced nitrogen abundances, relative to field stars at the same metallicity. These stars had not been previously recognized as a halo subpopulation, since there would typically be only one or two of them in any given halo study (e.g., \cite[Carretta et al. 2010]{CB10}). With the significantly larger number of halo stars with SEGUE spectra, it was much clearer that this small fraction forms a distinct subpopulation of halo stars.

A followup study (\cite[Martell et al. 2011]{MS11}) used the same technique to search the SDSS-III SEGUE-2 survey for globular cluster migrants in the halo field. A similar fraction of CN-strong and CH-weak stars were found as in \cite{MG10}. The observations in SEGUE-2 were deliberately targeted at more distant halo stars than in SEGUE-1, allowing for an evaluation of whether the fraction of globular cluster migrants changes as a function of Galactocentric distance, which might be expected because the majority of halo globular clusters orbit within 20 kpc. There is an indication that the fraction of migrant stars is higher in the inner region of the halo, but the number of stars involved at large distances is too small to make strong claims.

Other large-scale spectroscopic surveys have been searched for similar chemically taggable globular cluster stars in the halo. Using data from the Gaia-ESO Survey (\cite[Gilmore et al. 2012]{GR12}), \cite{LK15} identified one halo star with abundances of Mg and Al significantly different from the field stars but consistent with the characteristic globular cluster abundance pattern. While Mg and Al abundances do not always participate in the anticorrelated abundance pattern, this star appears to have escaped from a cluster where they did: it is well distinct from field stars in the Mg-Al plane, and consistent with the extended anticorrelations seen in some globular clusters. The authors also compared the star's orbit to those of several globular clusters, and found that it was most consistent with $\omega$ Centauri.

\cite{MS16} carried out a similar chemical tagging search in the SDSS-IV APOGEE survey, using nitrogen and aluminium abundances to identify halo giants with globular cluster-like abundances. These particular elements were chosen because the APOGEE DR12 abundances for them are valid and reliable across the metallicity and temperature range in the data set, and the two elements tend to be correlated in globular clusters. The authors identified five candidate globular cluster migrants out of 253 halo giants, a fraction consistent with earlier searches for similar stars.

Intriguingly, \cite{SZ17} found indications of a similar abundance-defined subpopulation in the Galactic bulge, using the same N-Al abundance plane. Taking APOGEE giants within 3~kpc of the Galactic centre, the authors found that a fraction of those stars at all metallicities have distinctly high N abundances and a tendency for high Al abundances. This suggests that the destruction of globular clusters might have played a significant role in the early construction of the bulge, but the metallicity distribution of the nitrogen-rich stars does not match either the nitrogen-normal bulge stars or the globular clusters that orbit within the disk and bulge. A more detailed study of the abundances patterns and kinematics of these potential bulge migrants would help to clarify their origins.

Within the GALAH survey (\cite[De Silva et al. 2015]{DS15}), a chemical tagging search of halo giants is returning a similar result. Although the survey volume of GALAH is somewhat smaller than APOGEE (most dwarfs within 3~kpc, most giants within 7~kpc; \cite[Martell et al. 2017]{MS17}), samples of halo stars can be identified based on their height above the plane and orbital properties. Within a sample of 132 metal-poor ([Fe/H]$<-1.5$) giants located at least 5~kpc off the Galactic plane, Ho \& Martell (in prep) find that a small fraction are enriched in both sodium and aluminium. This work will be extended by expanding the set of elements used for chemical tagging and by expanding the set of halo stars to include Solar neighbourhood stars with halo-like orbits using proper motions from the \textit{Gaia} TGAS data (\cite[Michalik et al. 2015]{TGAS}; \cite[Lindegren et al. 2016]{GDR1}).

\section{Globular cluster stars in the halo field}\label{sec:interpretation}
The fraction of stars in a globular cluster that exhibit the anticorrelated abundance pattern is directly related to the cluster's present-day mass (\cite[Milone et al. 2017]{MP17}), which bears an inexact correspondence to its original mass, and the overall average fraction is roughly 50\%. The ordinary mass loss processes that liberate chemically taggable globular cluster migrants from clusters into the halo should be unaware of the stars' chemical compositions, implying that an equal number of chemically untaggable stars have also been transferred into the halo field through those same processes. Whether this is simply a result of mass loss from clusters that are still bound or a result of total dissolution of some number of clusters from the initial Galactic population is unclear. Regardless, if $2.5\%$ of halo stars are chemically taggable to globular clusters, and each chemically taggable cluster migrant has a chemically ordinary partner, then there are as many globular cluster stars in the halo field as there are currently in clusters.

The process by which globular clusters formed in the early Universe, forming at least $10^4~M_{\odot}$ in stars within a few hundred million years in high-density but dark matter-free subhalos and imprinting them with a universal but apparently unique light-element abundance pattern, is not yet clear. Models that posit the abundance pattern as a result of feedback between multiple generations of stars (e.g., \cite[Decressin et al. 2007]{DC07}, \cite[D'Ercole et al. 2010]{DD10}) require that the first generation contains a very large number of stars. If those chemically untaggable stars have also been shed into the halo, then the fraction of halo stars originating in globular clusters could be significantly higher than 5\%. Indeed, a simplistic requirement that globular clusters not produce more stars than are currently found in the halo restricts the initial mass of the first generation in these models to at most a factor of 18 more massive than at the present day, and a more complete consideration of the population properties of halo field stars and globular cluster stars restricts that factor even further.

Much work remains to build a more complete understanding of the role that globular clusters have played in the assembly of the Galactic halo. Observationally, a full chemical inventory of the stars identified so far as globular cluster migrants is crucial to confirming their origins within globular clusters and potentially identifying their clusters of origin. Observations of globular cluster formation \textit{in situ} at high redshift may be possible in the JWST era, which would be extremely helpful to the efforts to understand how globular clusters form and how they survive through the process of hierarchical galaxy formation to the present day. Theoretically, the most important development will be testable predictions from multiple cosmologically situated models for globular cluster formation. This will allow us to use the significant existing library of kinematic and abundance data on globular clusters and their stars to evaluate those models and ideally identify the true formation mechanism for globular clusters.

\end{document}